\pdfoutput=1
\documentclass[11pt,twoside,a4paper,cmspaper,final,collab]{cms-tdr}
\def\svnVersion{27430M}\def\cmsCernNo{2010-054}\def\cmsCernDate{\today}\def\cmsMessage{Submitted to Physical Review Letters}

\begin{document}\cmsNoteHeader{EXO-10-007}
\hyphenation{env-iron-men-tal}
\hyphenation{had-ron-i-za-tion}
\hyphenation{cal-or-i-me-ter}
\hyphenation{de-vices}
\RCS$Revision: 27357 $
\RCS$HeadURL: svn+ssh://alverson@svn.cern.ch/reps/tdr2/papers/EXO-10-007/trunk/EXO-10-007.tex $
\RCS$Id: EXO-10-007.tex 27357 2010-12-17 04:31:47Z darinb $
%
%
%

\providecommand {\etal}{\mbox{et al.}\xspace} 
\providecommand {\ie}{\mbox{i.e.}\xspace}     
\providecommand {\eg}{\mbox{e.g.}\xspace}     
\providecommand {\etc}{\mbox{etc.}\xspace}     
\providecommand {\vs}{\mbox{\sl vs.}\xspace}      
\providecommand {\mdash}{\ensuremath{\mathrm{-}}} 

\providecommand {\Lone}{Level-1\xspace} 
\providecommand {\Ltwo}{Level-2\xspace}
\providecommand {\Lthree}{Level-3\xspace}

\providecommand{\ACERMC} {\textsc{AcerMC}\xspace}
\providecommand{\ALPGEN} {{\textsc{alpgen}}\xspace}
\providecommand{\CHARYBDIS} {{\textsc{charybdis}}\xspace}
\providecommand{\CMKIN} {\textsc{cmkin}\xspace}
\providecommand{\CMSIM} {{\textsc{cmsim}}\xspace}
\providecommand{\CMSSW} {{\textsc{cmssw}}\xspace}
\providecommand{\COBRA} {{\textsc{cobra}}\xspace}
\providecommand{\COCOA} {{\textsc{cocoa}}\xspace}
\providecommand{\COMPHEP} {\textsc{CompHEP}\xspace}
\providecommand{\EVTGEN} {{\textsc{evtgen}}\xspace}
\providecommand{\FAMOS} {{\textsc{famos}}\xspace}
\providecommand{\GARCON} {\textsc{garcon}\xspace}
\providecommand{\GARFIELD} {{\textsc{garfield}}\xspace}
\providecommand{\GEANE} {{\textsc{geane}}\xspace}
\providecommand{\GEANTfour} {{\textsc{geant4}}\xspace}
\providecommand{\GEANTthree} {{\textsc{geant3}}\xspace}
\providecommand{\GEANT} {{\textsc{geant}}\xspace}
\providecommand{\HDECAY} {\textsc{hdecay}\xspace}
\providecommand{\HERWIG} {{\textsc{herwig}}\xspace}
\providecommand{\HIGLU} {{\textsc{higlu}}\xspace}
\providecommand{\HIJING} {{\textsc{hijing}}\xspace}
\providecommand{\IGUANA} {\textsc{iguana}\xspace}
\providecommand{\ISAJET} {{\textsc{isajet}}\xspace}
\providecommand{\ISAPYTHIA} {{\textsc{isapythia}}\xspace}
\providecommand{\ISASUGRA} {{\textsc{isasugra}}\xspace}
\providecommand{\ISASUSY} {{\textsc{isasusy}}\xspace}
\providecommand{\ISAWIG} {{\textsc{isawig}}\xspace}
\providecommand{\MADGRAPH} {\textsc{MadGraph}\xspace}
\providecommand{\MCATNLO} {\textsc{mc@nlo}\xspace}
\providecommand{\MCFM} {\textsc{mcfm}\xspace}
\providecommand{\MILLEPEDE} {{\textsc{millepede}}\xspace}
\providecommand{\ORCA} {{\textsc{orca}}\xspace}
\providecommand{\OSCAR} {{\textsc{oscar}}\xspace}
\providecommand{\PHOTOS} {\textsc{photos}\xspace}
\providecommand{\PROSPINO} {\textsc{prospino}\xspace}
\providecommand{\PYTHIA} {{\textsc{pythia}}\xspace}
\providecommand{\SHERPA} {{\textsc{sherpa}}\xspace}
\providecommand{\TAUOLA} {\textsc{tauola}\xspace}
\providecommand{\TOPREX} {\textsc{TopReX}\xspace}
\providecommand{\XDAQ} {{\textsc{xdaq}}\xspace}

\providecommand {\DZERO}{D\O\xspace}     


\providecommand{\de}{\ensuremath{^\circ}}
\providecommand{\ten}[1]{\ensuremath{\times \text{10}^\text{#1}}}
\providecommand{\unit}[1]{\ensuremath{\text{\,#1}}\xspace}
\providecommand{\mum}{\ensuremath{\,\mu\text{m}}\xspace}
\providecommand{\micron}{\ensuremath{\,\mu\text{m}}\xspace}
\providecommand{\cm}{\ensuremath{\,\text{cm}}\xspace}
\providecommand{\mm}{\ensuremath{\,\text{mm}}\xspace}
\providecommand{\mus}{\ensuremath{\,\mu\text{s}}\xspace}
\providecommand{\keV}{\ensuremath{\,\text{ke\hspace{-.08em}V}}\xspace}
\providecommand{\MeV}{\ensuremath{\,\text{Me\hspace{-.08em}V}}\xspace}
\providecommand{\GeV}{\ensuremath{\,\text{Ge\hspace{-.08em}V}}\xspace}
\providecommand{\gev}{\GeV}
\providecommand{\TeV}{\ensuremath{\,\text{Te\hspace{-.08em}V}}\xspace}
\providecommand{\PeV}{\ensuremath{\,\text{Pe\hspace{-.08em}V}}\xspace}
\providecommand{\keVc}{\ensuremath{{\,\text{ke\hspace{-.08em}V\hspace{-0.16em}/\hspace{-0.08em}}c}}\xspace}
\providecommand{\MeVc}{\ensuremath{{\,\text{Me\hspace{-.08em}V\hspace{-0.16em}/\hspace{-0.08em}}c}}\xspace}
\providecommand{\GeVc}{\ensuremath{{\,\text{Ge\hspace{-.08em}V\hspace{-0.16em}/\hspace{-0.08em}}c}}\xspace}
\providecommand{\TeVc}{\ensuremath{{\,\text{Te\hspace{-.08em}V\hspace{-0.16em}/\hspace{-0.08em}}c}}\xspace}
\providecommand{\keVcc}{\ensuremath{{\,\text{ke\hspace{-.08em}V\hspace{-0.16em}/\hspace{-0.08em}}c^\text{2}}}\xspace}
\providecommand{\MeVcc}{\ensuremath{{\,\text{Me\hspace{-.08em}V\hspace{-0.16em}/\hspace{-0.08em}}c^\text{2}}}\xspace}
\providecommand{\GeVcc}{\ensuremath{{\,\text{Ge\hspace{-.08em}V\hspace{-0.16em}/\hspace{-0.08em}}c^\text{2}}}\xspace}
\providecommand{\TeVcc}{\ensuremath{{\,\text{Te\hspace{-.08em}V\hspace{-0.16em}/\hspace{-0.08em}}c^\text{2}}}\xspace}

\providecommand{\pbinv} {\mbox{\ensuremath{\,\text{pb}^\text{$-$1}}}\xspace}
\providecommand{\fbinv} {\mbox{\ensuremath{\,\text{fb}^\text{$-$1}}}\xspace}
\providecommand{\nbinv} {\mbox{\ensuremath{\,\text{nb}^\text{$-$1}}}\xspace}
\providecommand{\percms}{\ensuremath{\,\text{cm}^\text{$-$2}\,\text{s}^\text{$-$1}}\xspace}
\providecommand{\lumi}{\ensuremath{\mathcal{L}}\xspace}
\providecommand{\Lumi}{\ensuremath{\mathcal{L}}\xspace}
%
\providecommand{\LvLow}  {\ensuremath{\mathcal{L}=\text{10}^\text{32}\,\text{cm}^\text{$-$2}\,\text{s}^\text{$-$1}}\xspace}
\providecommand{\LLow}   {\ensuremath{\mathcal{L}=\text{10}^\text{33}\,\text{cm}^\text{$-$2}\,\text{s}^\text{$-$1}}\xspace}
\providecommand{\lowlumi}{\ensuremath{\mathcal{L}=\text{2}\times \text{10}^\text{33}\,\text{cm}^\text{$-$2}\,\text{s}^\text{$-$1}}\xspace}
\providecommand{\LMed}   {\ensuremath{\mathcal{L}=\text{2}\times \text{10}^\text{33}\,\text{cm}^\text{$-$2}\,\text{s}^\text{$-$1}}\xspace}
\providecommand{\LHigh}  {\ensuremath{\mathcal{L}=\text{10}^\text{34}\,\text{cm}^\text{$-$2}\,\text{s}^\text{$-$1}}\xspace}
\providecommand{\hilumi} {\ensuremath{\mathcal{L}=\text{10}^\text{34}\,\text{cm}^\text{$-$2}\,\text{s}^\text{$-$1}}\xspace}


\providecommand{\PT}{\ensuremath{p_{\mathrm{T}}}\xspace}
\providecommand{\pt}{\ensuremath{p_{\mathrm{T}}}\xspace}
\providecommand{\ET}{\ensuremath{E_{\mathrm{T}}}\xspace}
\providecommand{\HT}{\ensuremath{H_{\mathrm{T}}}\xspace}
\providecommand{\et}{\ensuremath{E_{\mathrm{T}}}\xspace}
\providecommand{\Em}{\ensuremath{E\hspace{-0.6em}/}\xspace}
\providecommand{\Pm}{\ensuremath{p\hspace{-0.5em}/}\xspace}
\providecommand{\PTm}{\ensuremath{{p}_\mathrm{T}\hspace{-1.02em}/}\xspace}
\providecommand{\PTslash}{\ensuremath{{p}_\mathrm{T}\hspace{-1.02em}/}\xspace}
\providecommand{\ETm}{\ensuremath{E_{\mathrm{T}}^{\text{miss}}}\xspace}
\providecommand{\ETslash}{\ensuremath{E_{\mathrm{T}}\hspace{-1.1em}/}\xspace}
\providecommand{\MET}{\ensuremath{E_{\mathrm{T}}^{\text{miss}}}\xspace}
\providecommand{\ETmiss}{\ensuremath{E_{\mathrm{T}}^{\text{miss}}}\xspace}
\providecommand{\VEtmiss}{\ensuremath{{\vec E}_{\mathrm{T}}^{\text{miss}}}\xspace}

\providecommand{\dd}[2]{\ensuremath{\frac{\mathrm{d} #1}{\mathrm{d} #2}}}

\ifthenelse{\boolean{cms@italic}}{\newcommand{\cmsSymbolFace}{\relax}}{\newcommand{\cmsSymbolFace}{\mathrm}}

\providecommand{\zp}{\ensuremath{\cmsSymbolFace{Z}^\prime}\xspace}
\providecommand{\JPsi}{\ensuremath{\cmsSymbolFace{J}\hspace{-.08em}/\hspace{-.14em}\psi}\xspace}
\providecommand{\Z}{\ensuremath{\cmsSymbolFace{Z}}\xspace}
\providecommand{\ttbar}{\ensuremath{\cmsSymbolFace{t}\overline{\cmsSymbolFace{t}}}\xspace}


\providecommand{\AFB}{\ensuremath{A_\text{FB}}\xspace}
\providecommand{\wangle}{\ensuremath{\sin^{2}\theta_{\text{eff}}^\text{lept}(M^2_\Z)}\xspace}
\providecommand{\stat}{\ensuremath{\,\text{(stat.)}}\xspace}
\providecommand{\syst}{\ensuremath{\,\text{(syst.)}}\xspace}
\providecommand{\kt}{\ensuremath{k_{\mathrm{T}}}\xspace}

\providecommand{\BC}{\ensuremath{\mathrm{B_{c}}}\xspace}
\providecommand{\bbarc}{\ensuremath{\mathrm{\overline{b}c}}\xspace}
\providecommand{\bbbar}{\ensuremath{\mathrm{b\overline{b}}}\xspace}
\providecommand{\ccbar}{\ensuremath{\mathrm{c\overline{c}}}\xspace}
\providecommand{\bspsiphi}{\ensuremath{\mathrm{B_s} \to \JPsi\, \phi}\xspace}
\providecommand{\EE}{\ensuremath{\mathrm{e^+e^-}}\xspace}
\providecommand{\MM}{\ensuremath{\mu^+\mu^-}\xspace}
\providecommand{\TT}{\ensuremath{\tau^+\tau^-}\xspace}

\providecommand{\HGG}{\ensuremath{\mathrm{H}\to\gamma\gamma}}
\providecommand{\GAMJET}{\ensuremath{\gamma + \text{jet}}}
\providecommand{\PPTOJETS}{\ensuremath{\mathrm{pp}\to\text{jets}}}
\providecommand{\PPTOGG}{\ensuremath{\mathrm{pp}\to\gamma\gamma}}
\providecommand{\PPTOGAMJET}{\ensuremath{\mathrm{pp}\to\gamma + \mathrm{jet}}}
\providecommand{\MH}{\ensuremath{M_{\mathrm{H}}}}
\providecommand{\RNINE}{\ensuremath{R_\mathrm{9}}}
\providecommand{\DR}{\ensuremath{\Delta R}}

%

\providecommand{\ga}{\ensuremath{\gtrsim}}
\providecommand{\la}{\ensuremath{\lesssim}}
\providecommand{\swsq}{\ensuremath{\sin^2\theta_\mathrm{W}}\xspace}
\providecommand{\cwsq}{\ensuremath{\cos^2\theta_\mathrm{W}}\xspace}
\providecommand{\tanb}{\ensuremath{\tan\beta}\xspace}
\providecommand{\tanbsq}{\ensuremath{\tan^{2}\beta}\xspace}
\providecommand{\sidb}{\ensuremath{\sin 2\beta}\xspace}
\providecommand{\alpS}{\ensuremath{\alpha_S}\xspace}
\providecommand{\alpt}{\ensuremath{\tilde{\alpha}}\xspace}

\providecommand{\QL}{\ensuremath{\mathrm{Q}_\mathrm{L}}\xspace}
\providecommand{\sQ}{\ensuremath{\tilde{\mathrm{Q}}}\xspace}
\providecommand{\sQL}{\ensuremath{\tilde{\mathrm{Q}}_\mathrm{L}}\xspace}
\providecommand{\ULC}{\ensuremath{\mathrm{U}_\mathrm{L}^\mathrm{C}}\xspace}
\providecommand{\sUC}{\ensuremath{\tilde{\mathrm{U}}^\mathrm{C}}\xspace}
\providecommand{\sULC}{\ensuremath{\tilde{\mathrm{U}}_\mathrm{L}^\mathrm{C}}\xspace}
\providecommand{\DLC}{\ensuremath{\mathrm{D}_\mathrm{L}^\mathrm{C}}\xspace}
\providecommand{\sDC}{\ensuremath{\tilde{\mathrm{D}}^\mathrm{C}}\xspace}
\providecommand{\sDLC}{\ensuremath{\tilde{\mathrm{D}}_\mathrm{L}^\mathrm{C}}\xspace}
\providecommand{\LL}{\ensuremath{\mathrm{L}_\mathrm{L}}\xspace}
\providecommand{\sL}{\ensuremath{\tilde{\mathrm{L}}}\xspace}
\providecommand{\sLL}{\ensuremath{\tilde{\mathrm{L}}_\mathrm{L}}\xspace}
\providecommand{\ELC}{\ensuremath{\mathrm{E}_\mathrm{L}^\mathrm{C}}\xspace}
\providecommand{\sEC}{\ensuremath{\tilde{\mathrm{E}}^\mathrm{C}}\xspace}
\providecommand{\sELC}{\ensuremath{\tilde{\mathrm{E}}_\mathrm{L}^\mathrm{C}}\xspace}
\providecommand{\sEL}{\ensuremath{\tilde{\mathrm{E}}_\mathrm{L}}\xspace}
\providecommand{\sER}{\ensuremath{\tilde{\mathrm{E}}_\mathrm{R}}\xspace}
\providecommand{\sFer}{\ensuremath{\tilde{\mathrm{f}}}\xspace}
\providecommand{\sQua}{\ensuremath{\tilde{\mathrm{q}}}\xspace}
\providecommand{\sUp}{\ensuremath{\tilde{\mathrm{u}}}\xspace}
\providecommand{\suL}{\ensuremath{\tilde{\mathrm{u}}_\mathrm{L}}\xspace}
\providecommand{\suR}{\ensuremath{\tilde{\mathrm{u}}_\mathrm{R}}\xspace}
\providecommand{\sDw}{\ensuremath{\tilde{\mathrm{d}}}\xspace}
\providecommand{\sdL}{\ensuremath{\tilde{\mathrm{d}}_\mathrm{L}}\xspace}
\providecommand{\sdR}{\ensuremath{\tilde{\mathrm{d}}_\mathrm{R}}\xspace}
\providecommand{\sTop}{\ensuremath{\tilde{\mathrm{t}}}\xspace}
\providecommand{\stL}{\ensuremath{\tilde{\mathrm{t}}_\mathrm{L}}\xspace}
\providecommand{\stR}{\ensuremath{\tilde{\mathrm{t}}_\mathrm{R}}\xspace}
\providecommand{\stone}{\ensuremath{\tilde{\mathrm{t}}_1}\xspace}
\providecommand{\sttwo}{\ensuremath{\tilde{\mathrm{t}}_2}\xspace}
\providecommand{\sBot}{\ensuremath{\tilde{\mathrm{b}}}\xspace}
\providecommand{\sbL}{\ensuremath{\tilde{\mathrm{b}}_\mathrm{L}}\xspace}
\providecommand{\sbR}{\ensuremath{\tilde{\mathrm{b}}_\mathrm{R}}\xspace}
\providecommand{\sbone}{\ensuremath{\tilde{\mathrm{b}}_1}\xspace}
\providecommand{\sbtwo}{\ensuremath{\tilde{\mathrm{b}}_2}\xspace}
\providecommand{\sLep}{\ensuremath{\tilde{\mathrm{l}}}\xspace}
\providecommand{\sLepC}{\ensuremath{\tilde{\mathrm{l}}^\mathrm{C}}\xspace}
\providecommand{\sEl}{\ensuremath{\tilde{\mathrm{e}}}\xspace}
\providecommand{\sElC}{\ensuremath{\tilde{\mathrm{e}}^\mathrm{C}}\xspace}
\providecommand{\seL}{\ensuremath{\tilde{\mathrm{e}}_\mathrm{L}}\xspace}
\providecommand{\seR}{\ensuremath{\tilde{\mathrm{e}}_\mathrm{R}}\xspace}
\providecommand{\snL}{\ensuremath{\tilde{\nu}_L}\xspace}
\providecommand{\sMu}{\ensuremath{\tilde{\mu}}\xspace}
\providecommand{\sNu}{\ensuremath{\tilde{\nu}}\xspace}
\providecommand{\sTau}{\ensuremath{\tilde{\tau}}\xspace}
\providecommand{\Glu}{\ensuremath{\mathrm{g}}\xspace}
\providecommand{\sGlu}{\ensuremath{\tilde{\mathrm{g}}}\xspace}
\providecommand{\Wpm}{\ensuremath{\mathrm{W}^{\pm}}\xspace}
\providecommand{\sWpm}{\ensuremath{\tilde{\mathrm{W}}^{\pm}}\xspace}
\providecommand{\Wz}{\ensuremath{\mathrm{W}^{0}}\xspace}
\providecommand{\sWz}{\ensuremath{\tilde{\mathrm{W}}^{0}}\xspace}
\providecommand{\sWino}{\ensuremath{\tilde{\mathrm{W}}}\xspace}
\providecommand{\Bz}{\ensuremath{\mathrm{B}^{0}}\xspace}
\providecommand{\sBz}{\ensuremath{\tilde{\mathrm{B}}^{0}}\xspace}
\providecommand{\sBino}{\ensuremath{\tilde{\mathrm{B}}}\xspace}
\providecommand{\Zz}{\ensuremath{\mathrm{Z}^{0}}\xspace}
\providecommand{\sZino}{\ensuremath{\tilde{\mathrm{Z}}^{0}}\xspace}
\providecommand{\sGam}{\ensuremath{\tilde{\gamma}}\xspace}
\providecommand{\chiz}{\ensuremath{\tilde{\chi}^{0}}\xspace}
\providecommand{\chip}{\ensuremath{\tilde{\chi}^{+}}\xspace}
\providecommand{\chim}{\ensuremath{\tilde{\chi}^{-}}\xspace}
\providecommand{\chipm}{\ensuremath{\tilde{\chi}^{\pm}}\xspace}
\providecommand{\Hone}{\ensuremath{\mathrm{H}_\mathrm{d}}\xspace}
\providecommand{\sHone}{\ensuremath{\tilde{\mathrm{H}}_\mathrm{d}}\xspace}
\providecommand{\Htwo}{\ensuremath{\mathrm{H}_\mathrm{u}}\xspace}
\providecommand{\sHtwo}{\ensuremath{\tilde{\mathrm{H}}_\mathrm{u}}\xspace}
\providecommand{\sHig}{\ensuremath{\tilde{\mathrm{H}}}\xspace}
\providecommand{\sHa}{\ensuremath{\tilde{\mathrm{H}}_\mathrm{a}}\xspace}
\providecommand{\sHb}{\ensuremath{\tilde{\mathrm{H}}_\mathrm{b}}\xspace}
\providecommand{\sHpm}{\ensuremath{\tilde{\mathrm{H}}^{\pm}}\xspace}
\providecommand{\hz}{\ensuremath{\mathrm{h}^{0}}\xspace}
\providecommand{\Hz}{\ensuremath{\mathrm{H}^{0}}\xspace}
\providecommand{\Az}{\ensuremath{\mathrm{A}^{0}}\xspace}
\providecommand{\Hpm}{\ensuremath{\mathrm{H}^{\pm}}\xspace}
\providecommand{\sGra}{\ensuremath{\tilde{\mathrm{G}}}\xspace}
\providecommand{\mtil}{\ensuremath{\tilde{m}}\xspace}
\providecommand{\rpv}{\ensuremath{\rlap{\kern.2em/}R}\xspace}
\providecommand{\LLE}{\ensuremath{LL\bar{E}}\xspace}
\providecommand{\LQD}{\ensuremath{LQ\bar{D}}\xspace}
\providecommand{\UDD}{\ensuremath{\overline{UDD}}\xspace}
\providecommand{\Lam}{\ensuremath{\lambda}\xspace}
\providecommand{\Lamp}{\ensuremath{\lambda'}\xspace}
\providecommand{\Lampp}{\ensuremath{\lambda''}\xspace}
\providecommand{\spinbd}[2]{\ensuremath{\bar{#1}_{\dot{#2}}}\xspace}

\providecommand{\MD}{\ensuremath{{M_\mathrm{D}}}\xspace}
\providecommand{\Mpl}{\ensuremath{{M_\mathrm{Pl}}}\xspace}
\providecommand{\Rinv} {\ensuremath{{R}^{-1}}\xspace} 

\title{ Search for Pair Production of Second-Generation Scalar Leptoquarks in $\Pp \Pp$ Collisions at
         $\sqrt{s}= 7$ TeV
       }            
\address[cern]{CERN}
\author[cern]{The CMS Collaboration}
\date{\today}
\abstract{
  A search for pair production of second-generation scalar leptoquarks in the final state with two
  muons and two jets is performed using proton-proton collision data at $\sqrt{s}= 7$~TeV collected
  by the CMS detector at the LHC. The data sample used corresponds to an integrated luminosity
  of 34~pb$^{-1}$. The number of observed events is in good agreement with the predictions
  from the standard model processes. An upper limit is set on the second-generation leptoquark cross
  section times $\beta^2$ as a function of the leptoquark mass, and leptoquarks with masses
  below 394~GeV are excluded at a 95\% confidence level for $\beta = 1$, where $\beta$ is the
  leptoquark branching fraction
  into a muon and a quark. These limits are the most stringent to date.
}
\hypersetup{%
             pdfauthor={CMS Collaboration},%
             pdftitle    = {Search for Pair Production of Second-Generation Scalar Leptoquarks in
                            pp Collisions at sqrt(s) = 7 TeV},
             pdfsubject  = {CMS},%
             pdfkeywords = {CMS, physics}
            }

\maketitle 
Several extensions of the standard model ~\cite{mBRW,techni1,techni2,techni3,superstring} predict
the existence of leptoquarks (LQ), hypothetical particles that carry both lepton and baryon numbers
and couple to both leptons and quarks. Leptoquarks are fractionally charged and can be either scalar
or vector particles. In order to satisfy constraints from flavour-changing neutral currents and rare
pion and kaon decays~\cite{fcnc,Buchmuller:1986zs}, leptoquarks are restricted to couple to a single
lepton-quark generation.

In proton-proton collisions at the CERN Large Hadron Collider (LHC) the dominant mechanisms for pair
production of scalar leptoquarks are gluon-gluon fusion and $q\bar{q}$-annihilation. The cross
section depends on the strong coupling constant and the LQ mass and has been calculated at
Next-to-Leading-Order (NLO) ~\cite{PhysRevD.71.057503}; the dependence on the Yukawa coupling
$\lambda$ is negligible~\cite{PhysRevD.71.057503}. Leptoquarks decay to a quark
and a charged lepton of the same generation with unknown branching fraction $\beta$ and to a quark
and a neutrino with branching fraction $(1-\beta)$. In this analysis, we consider the decay of a
second-generation leptoquark to a muon and a quark.

Several experiments have searched for leptoquarks, but so far no evidence has been observed. A
review of LQ phenomenology and searches can be found in ~\cite{Acosta, Hewett}. The most recent
limits from the \DZERO experiment at the Fermilab Tevatron collider exclude
second-generation scalar leptoquarks with masses below 316~GeV for $\beta = 1$, based on
proton-antiproton collisions at $\sqrt{s}= 1.96$ TeV~\cite{D0} .

This Letter describes a search for pair production of second-generation scalar leptoquarks with
the CMS experiment using LHC proton-proton collisions at $\sqrt{s}=7$~TeV. The data sample used
corresponds to an integrated luminosity of $34.0 \pm 3.7 \, \text{pb}^{-1}$.

The CMS detector, described in detail elsewhere~\cite{CMS}, uses a cylindrical coordinate system
with the $z$ axis along the counterclockwise beam direction 
{  The angles $\theta$ and $\phi$
are the polar and azimuthal angles, respectively.  Pseudorapidity is defined as $\eta =
-\ln[\tan(\theta/2)]$, where $\theta$ is measured with respect to the $+z$-axis.}
The central feature of the CMS apparatus is a superconducting solenoid,
of 6~m internal diameter, providing a field of 3.8~T. Within the field volume are the silicon
pixel and strip tracker, the crystal electromagnetic calorimeter (ECAL) and the brass/scintillator
hadron calorimeter (HCAL). Muons are measured in gas-ionization detectors embedded in the steel
return yoke. In addition to the barrel and endcap detectors, CMS has extensive forward calorimetry.
 The inner tracking system consists of a silicon pixel and strip
tracker, providing the required granularity and precision for the reconstruction of vertices of
charged particles having pseudorapidities $|\eta|<2.5$. The ECAL and HCAL are used to measure the
energies of photons, electrons,
and hadrons within a region of $|\eta|<3.0$. The three muon systems surrounding the solenoid cover
a region $|\eta|<2.4$ and are composed of drift tubes in the barrel region $(|\eta|<1.2)$, of
cathode strip chambers in the endcaps $(0.9<|\eta|<2.4)$, and of resistive plate chambers in both
the barrel region and the endcaps $(|\eta|<1.6)$. Events are recorded based on a first-level
trigger decision coming from either the calorimeter or muon systems. The final trigger decision is
 based on the information from all subsystems, which is passed
on to the High Level Trigger (HLT), consisting of a farm of computers running a version of the
reconstruction software optimized for fast processing.
The signature of the decay of pair-produced second-generation leptoquarks studied here consists of
two muons and two jets with high transverse momentum ($p_T$).
Events are selected by a single muon trigger without isolation requirements and with lower $p_T$
thresholds dependent upon the instantaneous luminosity. The HLT efficiency for events
with two high $p_T$ muons is always greater than 99\%.

The Monte Carlo (MC) signal events are generated using the {\PYTHIA}~\cite{Sjostrand:2006za} generator
(version 6.422) in the LQ mass range 250-600~GeV. The CTEQ6.6~\cite{Nadolsky:2008zw}
Parton Distribution Function (PDF) sets are used.
The main background processes that can mimic the signature of the LQ signal are
$\Z/\gamma^* + \text{jets}$, \ttbar, $VV$ $(\PW\PW, \Z\Z, \PW\Z)$, $\PW + \text{jets}$, and multijet events.
The \ttbar, $VV$, and muon-enriched multi-jets events are generated with
{\MADGRAPH}~\cite{MADGRAPH1,MADGRAPH2}; $\Z/\gamma^{*} + \text{jets}$ and $\PW + \text{jets}$ events
are generated with {\sc alpgen}. In {\MADGRAPH} and {\ALPGEN} samples, parton showering
and hadronization is performed with {\PYTHIA}.

Muons are reconstructed as tracks in the muon system that are matched to the tracks reconstructed
in the inner tracking system. Muons are required to have $p_T > 30$~GeV, $|\eta|<2.4$. The muon
relative isolation paramter is defined as the scalar sum of the $p_T$ of all tracks in the
tracker and the transverse energies of hits in
the ECAL and HCAL in a cone of $\Delta R = \sqrt{(\Delta\eta)^2 + (\Delta\phi)^2} = 0.3$
around the muon track, excluding the contribution from the muon itself,
divided by the muon $p_T$. Muons are required to have a  relative isolation parameter less than 0.05.
 $\Delta\eta$ and $\Delta\phi$ are the pseudorapidity and
azimuthal angle differences between the muon track and other reconstructed tracks or hits in the
calorimeter. To have a precise measurement of the transverse impact parameter of the
muon relative to the beam-spot position, only muons with tracks containing
more than 10 hits in the silicon tracker are considered. To reject muons from cosmic rays, the
transverse impact parameter is required to be less than 2~mm. In addition, the two muon candidates
are required to be separated from each other by at least $\Delta R = 0.3$ and at least one muon
must be in the pseudorapidity region $|\eta|<2.1$. The efficiency of selecting dimuon events is
61-70\% for the LQ mass range of 200-500 GeV.

Jets are reconstructed using the anti-k$_\text{T}$~\cite{antikt} algorithm with a distance parameter
$R = 0.5$ and are required to have $p_T > 30$~GeV and $|\eta|<3.0$. Jet-energy-scale corrections
derived from MC simulated events are applied to establish a relative uniform response
in $\eta$ and an absolute uniform response in $p_T$. A residual jet energy correction is derived
from data by looking at the balance in $p_T$ in dijet events, and it is applied to jets in data.

Additional selection requirements are placed on two variables, which are effective at
discriminating the LQ signal from the major sources of background. The first is the dimuon invariant mass,
$M_{\mu\mu}$. The second variable, $S_T$, is defined as the sum of the magnitudes of the $p_T$ of
the two highest $p_T$  muons and the two highest $p_T$ jets.
 The two muons in the signal events come from the decays of two high-mass particles,
and they tend to form a large invariant mass. Thus, events are selected if
$M_{\mu\mu} > 115$~GeV. This helps to reduce the contribution from
$\Z/\gamma^*+\text{jets}$ processes, which is one of the largest backgrounds.
In addition, the LQ pair is expected to have a large $S_T$.
The lower threshold on $S_T$ is optimized for different
LQ mass hypotheses by using a Bayesian approach~\cite{bayes1, bayes} to minimize the expected upper
limit on the LQ cross section in the absence
of an observed signal. The $S_T$ cut helps to
further reduce background sources, most noticeably \ttbar. The optimal $S_T$ threshold values
for each mass hypothesis are given in Table~\ref{tab:yields}. While the LQ signal is expected to
peak in the mass distribution of the $\mu$-jet pairs, we find
that the $S_T$ variable gives sufficient power of discrimination in the range of LQ masses considered. The $\mu$-jet mass distribution would nevertheless be important to establish the signal in case an excess is observed.

The contribution from \ttbar is estimated with the MC sample, using uncertainties determined
from data~\cite{ttdilep}.
The contribution from $\PW + \text{jets}$  is negligible once the full event selection is applied.
The small contribution from $VV$ is estimated from MC.
The multijet background is found to be negligible using a control data
sample of same-sign dimuon events. The background from $\Z/\gamma^* + \text{jets}$ is determined by
comparing $\Z/\gamma^* + \text{jets}$ events from data and MC samples in two different regions: at
the $\Z$ boson peak, $80 < M_{\mu\mu} < 100$~GeV, and in the high-mass region,
$M_{\mu\mu} > 115$~GeV. In the low-mass region, the ratio of data to MC events ($R_L$) is
determined to be $R_L = 1.28 \pm 0.14$ after selecting two muons and two jets with $p_T>$ 30 GeV,
and a preliminary requirement of $S_T > $250 GeV. This rescaling factor is applied to the number of
$\Z/\gamma^* +\text{jets}$ MC events in the high-mass region after the full selection.

\begin{figure}[htbp]
  \begin{center}

    \includegraphics[width=0.45\textwidth]{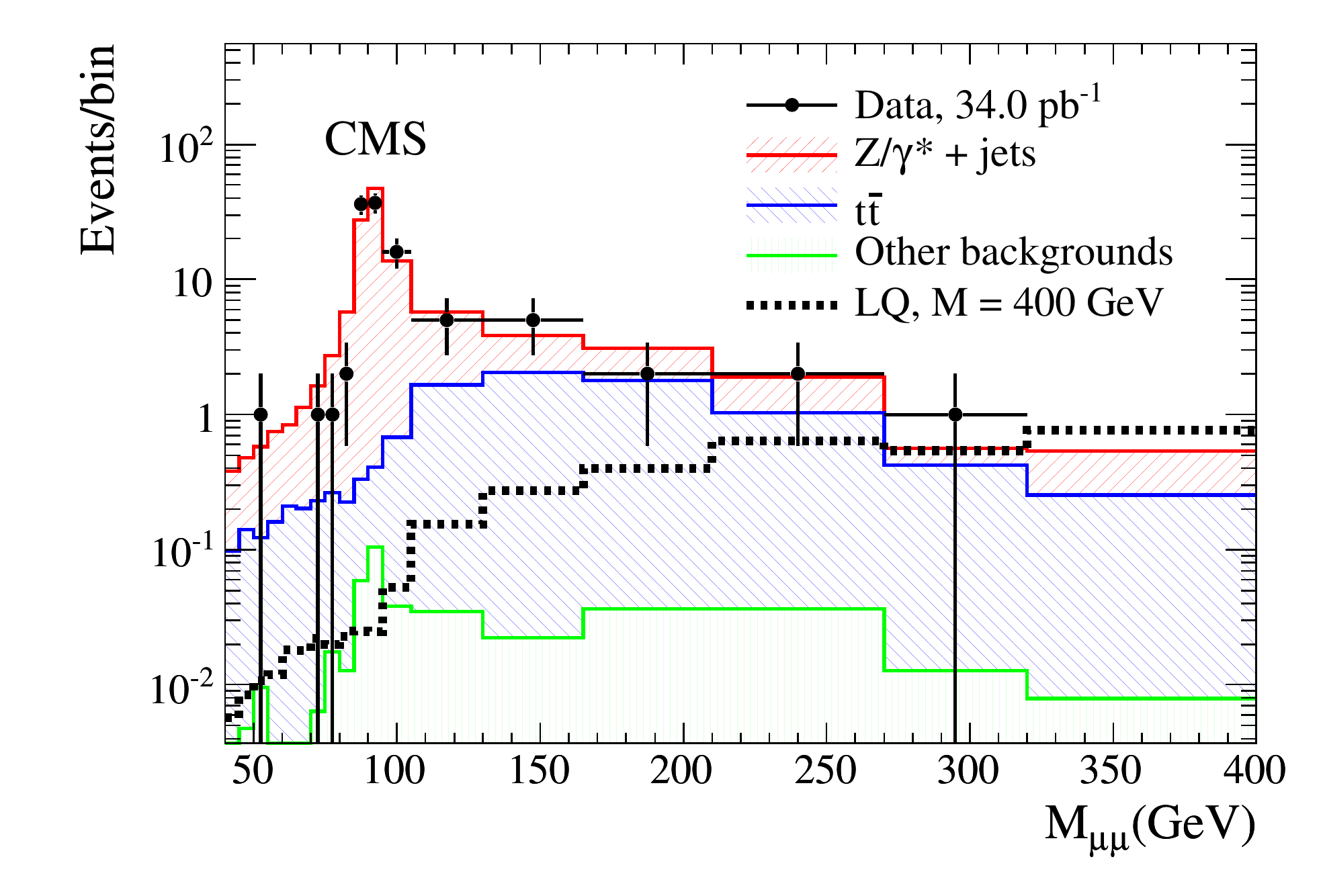}
    \includegraphics[width=0.45\textwidth]{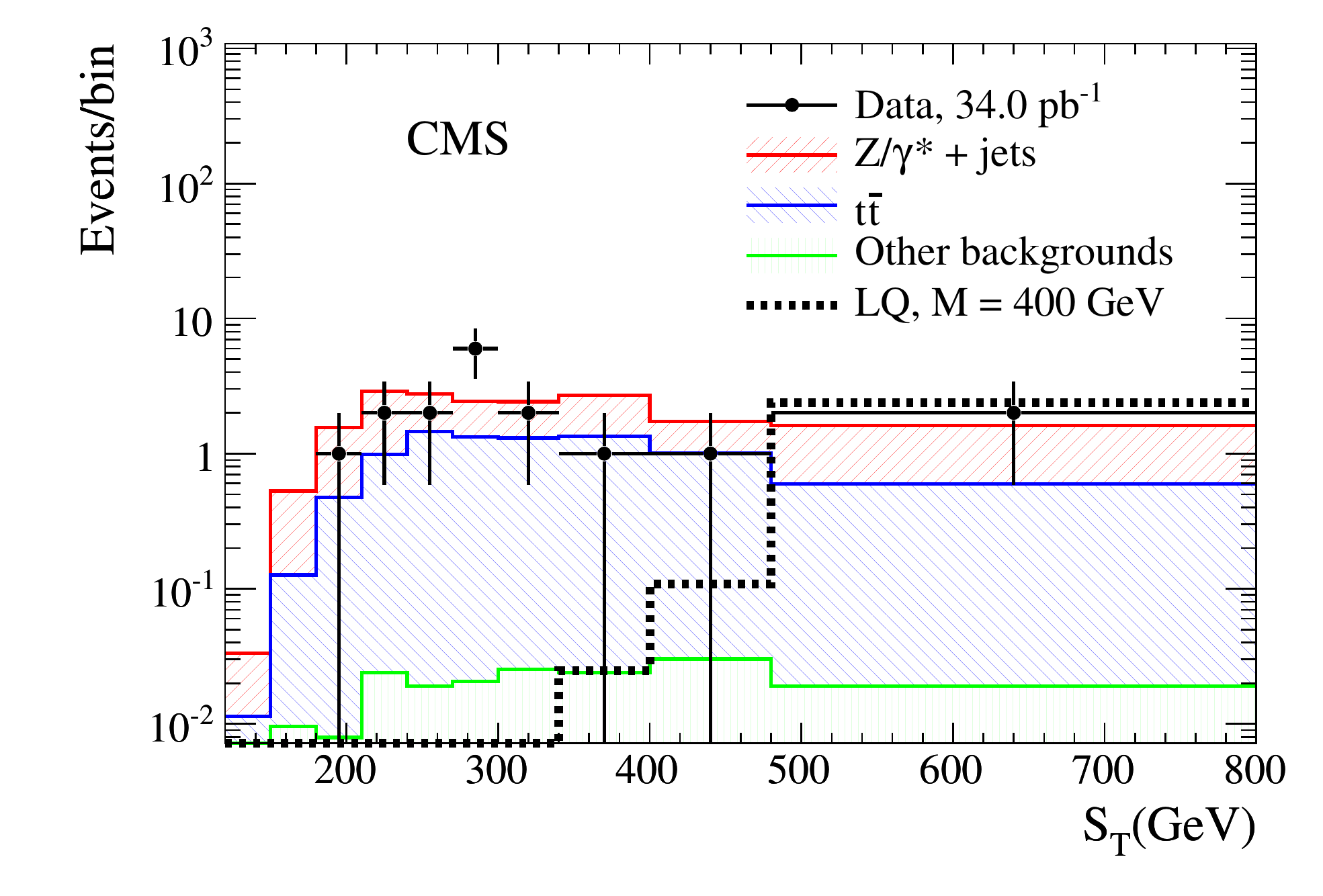}

    \caption{  The distribution of $M_{\mu \mu}$ (left) after requiring at least two muons and at
                least two jets with $p_T > 30$~GeV and $S_T > 250$ GeV, and the distribution of
                $S_T$ (right) after requiring at least two muons and at least two jets with
                $p_T > 30$~GeV and $M_{\mu\mu} > 115$~GeV. The $\Z/\gamma^* \to \mu\mu +
                \text{jets}$ contribution is rescaled by the factor $R_L = 1.28$. Other backgrounds
                correspond to $VV$, $\PW+\text{jets}$, and multijet processes. Uncertainties are
                statistical.
             }

    \label{fig:Mmm}

  \end{center}
\end{figure}

Reasonable agreement between data and MC predictions is observed at all selection levels. The
dimuon invariant mass is shown in Fig.~\ref{fig:Mmm} (left) after the initial selection of muons
and jets with $p_T>$ 30 GeV and a preliminary requirement of $S_T > 250$ GeV. The $M_{\mu \mu}$
distribution in data is consistent with the expected
SM background prediction. The $S_T$ distribution is
also shown in Fig.~\ref{fig:Mmm} (right) after the initial selection of muons and jets with
$p_T>$ 30 GeV and the additional requirement of $M_{\mu\mu} > 115$~GeV.

The event yields from data, expected LQ signal (for several mass hypotheses), signal selection
efficiency times acceptance, and expected standard model backgrounds are summarized in
Table~\ref{tab:yields}.

\begin{table*}[htbp]
  \begin{center}

    \caption{ The data event yields in 34.0 pb$^{-1}$ for different leptoquark mass hypotheses,
               together with the optimized $S_T$ threshold values (in GeV) for each mass, background
               predictions, number of expected LQ signal events (S), and signal selection efficiency
               times acceptance ($\epsilon_S$). M$_{LQ}$ and $S_T$ values are listed in GeV. The
               $\Z/\gamma^* \rightarrow \mu \mu+\text{jets}$ contribution is rescaled by the
               factor $R_L = 1.28$. Other backgrounds correspond to $VV$, $\PW+\text{jets}$, and
               multijet processes. Uncertainties are statistical.
             }

    \vspace{.25in}

    \scriptsize

    \begin{tabular}{| c|c c|c c c c|c|c|}\hline \hline
			$M_{LQ}$ & \multicolumn{2}{c|}{MC Signal Samples} & \multicolumn{4}{c|}{Monte Carlo Background Samples} & Events & Obs./Exp. \\
			($S_T$ Cut)  & Selected & Acceptance & \multicolumn{4}{c|}{Selected Events in} & in & {$95 \%$ C.L.} \\
      $\text{[GeV]}$  & Events & $\times$ Efficiency & \ttbar + jets & $\Z/\gamma^* +\text{jets}$ & Others & All & Data& u.l. on $\sigma $ [pb]\\ \hline
200 ($S_T>310$)&160$\pm$20&0.388$\pm$0.003&3.9$\pm$0.1&4.08$\pm$0.07&0.1$\pm$0.01&8.1$\pm$0.1&5&0.661/0.444\\
225 ($S_T>350$)&89$\pm$9&0.421$\pm$0.003&2.7$\pm$0.1&2.99$\pm$0.05&0.07$\pm$0.01&5.8$\pm$0.1&3&0.525/0.342\\
250 ($S_T>400$)&51$\pm$5&0.437$\pm$0.003&1.6$\pm$0.08&1.92$\pm$0.04&0.051$\pm$0.009&3.57$\pm$0.09&3&0.422/0.370\\
280 ($S_T>440$)&28$\pm$3&0.467$\pm$0.003&0.98$\pm$0.06&1.53$\pm$0.03&0.038$\pm$0.008&2.55$\pm$0.07&3&0.352/0.375\\
300 ($S_T>440$)&21$\pm$2&0.518$\pm$0.004&0.98$\pm$0.06&1.53$\pm$0.03&0.038$\pm$0.008&2.55$\pm$0.07&3&0.318/0.339\\
320 ($S_T>490$)&14$\pm$1&0.509$\pm$0.004&0.55$\pm$0.04&1.12$\pm$0.02&0.019$\pm$0.005&1.69$\pm$0.05&2&0.287/0.302\\
340 ($S_T>530$)&9$\pm$1&0.508$\pm$0.003&0.34$\pm$0.04&0.79$\pm$0.01&0.01$\pm$0.004&1.14$\pm$0.04&1&0.261/0.246\\
400 ($S_T>560$)&4.0$\pm$0.4&0.578$\pm$0.004&0.27$\pm$0.03&0.67$\pm$0.01&0.01$\pm$0.004&0.95$\pm$0.03&0&0.220/0.219\\
450 ($S_T>620$)&1.9$\pm$0.2&0.600$\pm$0.004&0.16$\pm$0.03&0.49$\pm$0.01&0.006$\pm$0.003&0.66$\pm$0.03&0&0.198/0.153\\
500 ($S_T>700$)&0.9$\pm$0.1&0.602$\pm$0.004&0.08$\pm$0.02&0.277$\pm$0.006&0.003$\pm$0.002&0.36$\pm$0.02&0&0.179/0.152\\\hline\hline

   \end{tabular}

   \label{tab:yields}

  \end{center}
\end{table*}

Several sources of systematic uncertainties are considered in this analysis. The uncertainty on the
integrated luminosity is taken as 11\%~\cite{lumi}. A 5\% systematic uncertainty is assigned to
the jet-energy-scale (JES) ~\cite{JESCorr} of each jet. A smaller, $\sim 1\%$ systematic
uncertainty comes from the muon momentum scale. The 300~GeV LQ signal
efficiency changes by 2\% and 1\% due to JES and muon momentum scale uncertainties, respectively.
The effect of the muon momentum scale uncertainty on the total background is estimated to be
$< 0.5\%$. The JES contributes 2\% to the estimate of the $\Z/\gamma^*+\text{jets}$ background
described above and 15\% to the estimate of the $VV$ background from MC. The statistical
uncertainty on the value of $R_L$ after a preselection requirement ($S_T>250$ GeV), 11\%,
is used as an uncertainty on the estimated $\Z/\gamma^*+\text{jets}$ background. A 41\% systematic
uncertainty is taken from the CMS	measurement of the \ttbar production cross section
~\cite{ttdilep} and assigned to the estimate of the \ttbar background; 
it includes the effect of JES on the estimate of the \ttbar background. The effect of jet energy
and muon momentum resolution on expected signal and backgrounds is found to be negligible. A 5\%
systematic uncertainty per muon is assigned due to differences in reconstruction, identification,
trigger, and isolation efficiencies between data and MC ~\cite{CMSWandZ}, resulting
in a 10\% uncertainty on the efficiency of selecting events with two muons both for the signal and
background processes. A theoretical uncertainty on
the LQ  signal production cross sections due to the choice of renormalization/factorization scales
 has been calculated by varying the scales between half
and twice the LQ mass, and is found to 14-15\% for LQ masses between 200 and 500 GeV.
 The 90\% C.L. PDF uncertainties on LQ cross section have been obtained
using the CTEQ6.6~\cite{Nadolsky:2008zw} PDF error set following a standard prescription 
and have been found to vary from 8 to 22 \% for leptoquarks in the mass range of
200 - 500 GeV ~\cite{PhysRevD.71.057503}. The effect of PDF uncertainties is less than 0.5\% on
signal acceptance. The PDF uncertainties are not considered for background sources with uncertainties
determined from data. The systematic uncertainties, their magnitude, and the relative impact on
the number of signal and background events are summarized in Table~\ref{tab:SysUncertainties}.

\begin{figure}[htb]
  \begin{center}

    \includegraphics[width=0.45\textwidth]{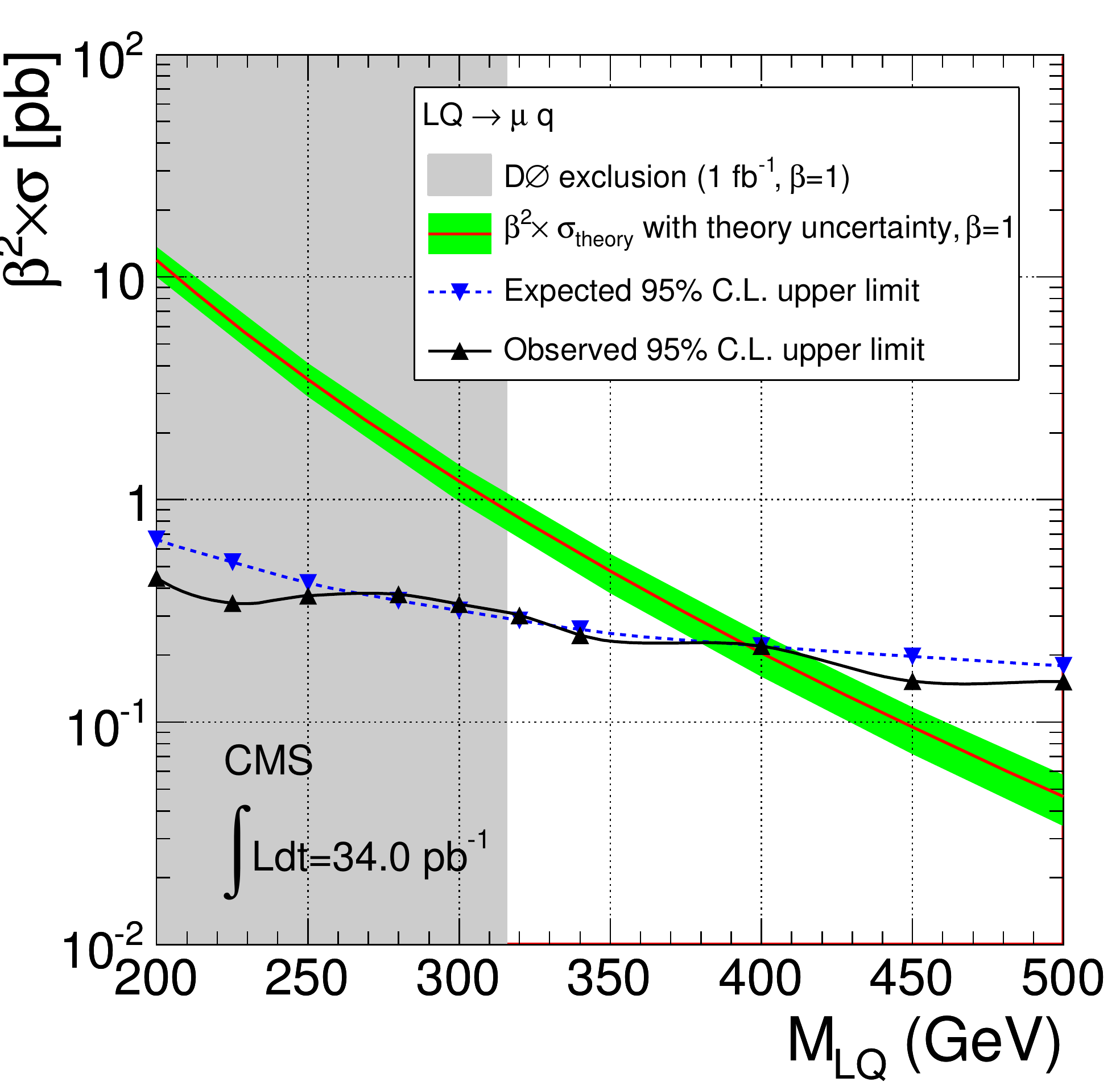}
    \includegraphics[width=0.45\textwidth]{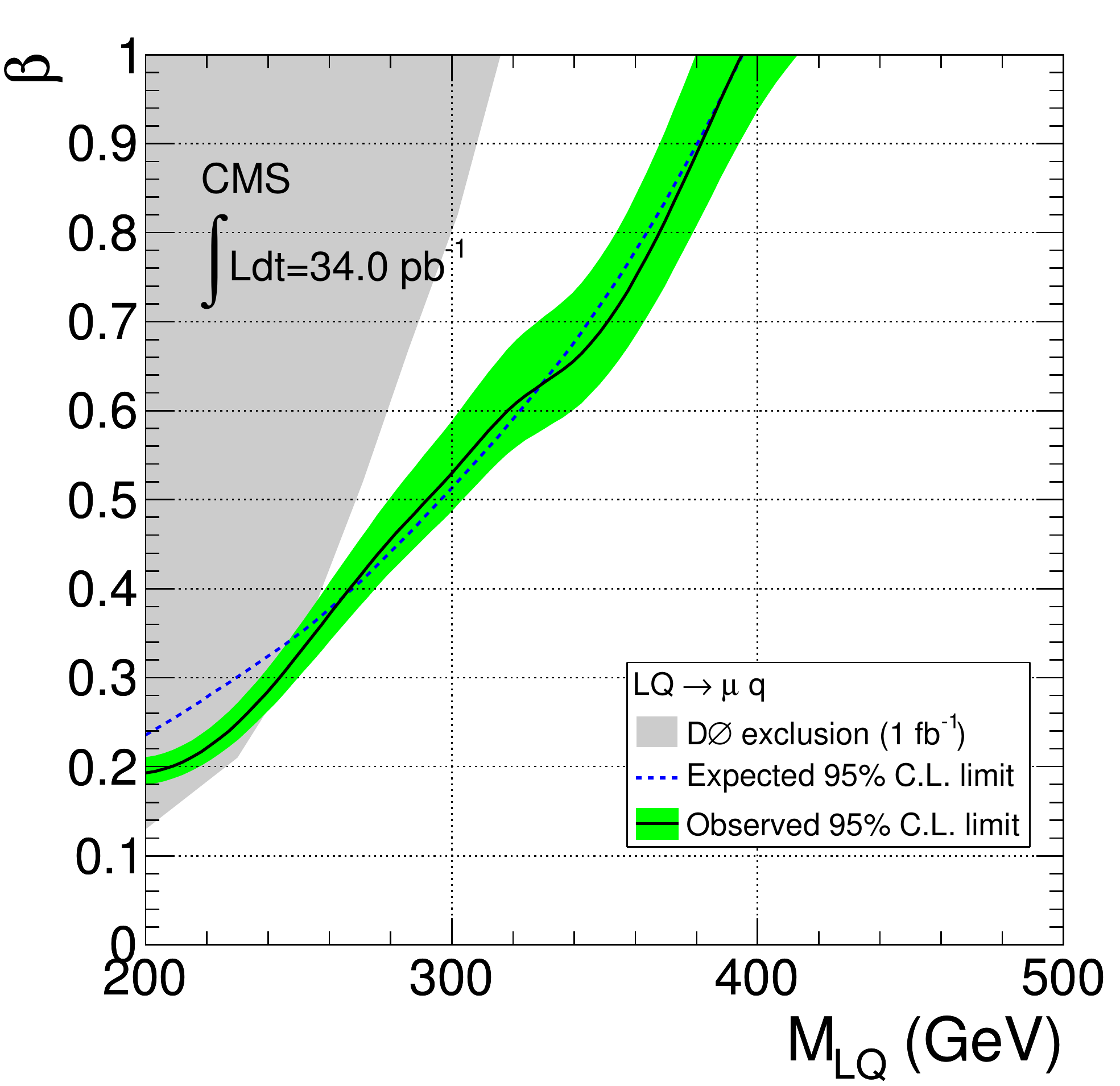}
    \caption{ (Left)
               The expected and observed 95\% C.L. upper limit on the scalar leptoquark pair
               production cross section multiplied by $\beta^2$ as a function of the LQ mass,
               together with the NLO theoretical cross section curve. The shaded band on the
               theoretical values includes CTEQ6.6 PDF uncertainties and the error on the leptoquark
               production cross section due to renormalization and factorization scale variation
               by a factor of two.   The shaded region
               is excluded by the current \DZERO limits~\cite{D0}.
               (Right)
               The minimum $\beta$ for 95\% C.L. exclusion of the leptoquark hypothesis
               as a function of leptoquark mass.
               The observed limit and corresponding uncertainty band is obtained by considering the
               observed upper limit and theoretical branching ratio and its uncertainty in
               the left-hand figure.
               Note: The shaded area excluded by the \DZERO experiment was
               determined with combined information from the decay channel with two muons and two
               jets and the decay channel with one muon, missing transverse energy, and two jets.
             }

    \label{fig:limit_plots2}
  \end{center}
\end{figure}

\begin{table}[htbp]
  \begin{center}

    \caption{ Systematic uncertainties and their effects on number of signal and background events }

    \vspace{.25in}

    \begin{tabular}{|l|c|c|c|}

      \hline

      Systematic                   & Magnitude   &  Effect on   &   Effect on        \\
      Uncertainty                  &             &  Signal      &  Background     \\ \hline \hline
      JES                          & $ 5\%$  &  $ 2\% $  &  $ -$              \\
      JES \& Data Backgr. Est.  & $ - $       &  $- $        &  $ 25\%$       \\
      Muon Momentum Scale          & $ 1\%$  &  $ 1\% $  &  $ < 0.5 \% $  \\
      Muon Pair Reco/ID/Iso        & $ 10\%$ &  $ 10\%$  &  $ < 0.05 \%$  \\
      Integrated Luminosity        & $ 11\%$ &  $ 11\%$  &  $ - $             \\
      \hline \hline
      Total                        &             & $ 15\%$   & $ 25\%$       \\
      \hline

    \end{tabular}

    \label{tab:SysUncertainties}
  \end{center}
\end{table}
\vspace*{0mm}
One candidate event survives the full selection criteria corresponding to a leptoquark mass
hypothesis of 340 GeV, and no candidates survive for criteria corresponding to masses greater than
450 GeV. An upper limit on the LQ cross section is set using a Bayesian method~\cite{bayes1, bayes}
with a flat signal prior. A log-normal probability density function is used to integrate over the
 systematic uncertainties. 
Using Poisson statistics, a 95\% confidence level (C.L.) upper limit is obtained on
$\sigma\times\beta^2$.
This is shown in Fig.~\ref{fig:limit_plots2} together with the NLO predictions for the scalar LQ
pair production cross section. The 95\% C.L. exclusion on $\beta$ as a function of LQ mass is
also shown in Fig.~\ref{fig:limit_plots2}. The systematic uncertainties reported in
Table~\ref{tab:SysUncertainties} are included in the calculation as nuisance parameters.
With the assumption that $\beta=1$, second-generation scalar leptoquarks with masses less than
394~GeV are excluded at 95\% C.L., 78 GeV higher than the limit set at the \DZERO Experiment at
the Tevatron~\cite{D0}. This is in agreement with the expected limit of 394~GeV. The corresponding
observed limit on cross section is 0.223 pb. If the lower edge of the
theoretical $\sigma \times \beta ^2$ curve is used, the observed (expected) limit on
LQ mass is 380 (378)~GeV and the observed limit on cross section is 0.225~pb.

\vspace*{0mm}
In summary, a search for pair production of second-generation scalar leptoquarks decaying to two
muons and two jets has been performed using 7 TeV $\Pp \Pp$ collision data corresponding to an integrated
luminosity of 34.0~pb$^{-1}$. The number of observed candidate events agrees well with the number
of expected standard model background events. A Bayesian approach that includes the treatment of
systematic uncertainties as nuisance parameters is used to set limits on the LQ cross section times
$\beta^2$ as a function of LQ mass. At 95\%~C.L., the pair production of second-generation scalar
leptoquarks with masses below 394~GeV is excluded for $\beta = 1$, where $\beta$ is the leptoquark
branching fraction into a muon and a quark. This is the most stringent limit to date on the
existence of second-generation scalar leptoquarks.

We extend our thanks to Michael  Kr\"amer for providing the tools for calculation of the
leptoquark theoretical cross section and PDF uncertainty.
We wish to congratulate our colleagues in the CERN accelerator departments for the excellent
performance of the LHC machine. We thank the technical and administrative staff at CERN and other
CMS institutes, and acknowledge support from: FMSR (Austria); FNRS and FWO (Belgium); CNPq, CAPES,
FAPERJ, and FAPESP (Brazil); MES (Bulgaria); CERN; CAS, MoST, and NSFC (China); COLCIENCIAS
(Colombia); MSES (Croatia); RPF (Cyprus); Academy of Sciences and NICPB (Estonia); Academy of
Finland, ME, and HIP (Finland); CEA and CNRS/IN2P3 (France); BMBF, DFG, and HGF (Germany); GSRT
(Greece); OTKA and NKTH (Hungary); DAE and DST (India); IPM (Iran); SFI (Ireland); INFN (Italy);
NRF and WCU (Korea); LAS (Lithuania); CINVESTAV, CONACYT, SEP, and UASLP-FAI (Mexico); PAEC
(Pakistan); SCSR (Poland); FCT (Portugal); JINR (Armenia, Belarus, Georgia, Ukraine, Uzbekistan);
MST and MAE (Russia); MSTD (Serbia); MICINN and CPAN (Spain); Swiss Funding Agencies (Switzerland);
NSC (Taipei); TUBITAK and TAEK (Turkey); STFC (United Kingdom); DOE and NSF (USA).
\clearpage
\bibliography{auto_generated}

\providecommand{\href}[2]{#2}\begingroup\raggedright\begin{thebibliography}{10}

\bibitem{mBRW}
W.~Buchm{\"u}ller {et~al.}, ``Leptoquarks in Lepton-Quark Collisions'',
  \textit{ Phys. Lett.} \textbf{ B191} (1987) 442.
\href{http://dx.doi.org/10.1016/0370-2693(87)90637-X}{\texttt{
  doi:10.1016/0370-2693(87)90637-X}}.

\bibitem{techni1}
S.~Dimopoulos and L.~Susskind, ``Mass Without Scalars'', \textit{ Nucl. Phys.}
  \textbf{ B155} (1979) 237.
\href{http://dx.doi.org/10.1016/0550-3213(79)90364-X}{\texttt{
  doi:10.1016/0550-3213(79)90364-X}}.

\bibitem{techni2}
S.~Dimopoulos, ``Technicolored signatures'', \textit{ Nucl. Phys.} \textbf{
  B168} (1980) 69.
\href{http://dx.doi.org/10.1016/0550-3213(80)90277-1}{\texttt{
  doi:10.1016/0550-3213(80)90277-1}}.

\bibitem{techni3}
E.~Eichten and K.~Lane, ``Dynamical Breaking of the Weak Interaction
  Symmetries'', \textit{ Phys. Lett.} \textbf{ B90} (1980) 125.
\href{http://dx.doi.org/10.1016/0370-2693(80)90065-9}{\texttt{
  doi:10.1016/0370-2693(80)90065-9}}.

\bibitem{superstring}
V.~Angelopoulos {et~al.}, ``Search for New Quarks Suggested by the
  Superstring'', \textit{ Nucl. Phys.} \textbf{ B292} (1987) 59.
\href{http://dx.doi.org/10.1016/0550-3213(87)90637-7}{\texttt{
  doi:10.1016/0550-3213(87)90637-7}}.

\bibitem{fcnc}
O.~Shanker, ``$\pi\ell2$, $K\ell3$ and $K^0-\bar{K^0}$ constraints on
  leptoquarks and supersymmetric particles'', \textit{ Nucl. Phys.} \textbf{
  B204} (1982) 375.
\href{http://dx.doi.org/10.1016/0550-3213(82)90196-1}{\texttt{
  doi:10.1016/0550-3213(82)90196-1}}.

\bibitem{Buchmuller:1986zs}
W.~Buchm{\"u}ller and D.~Wyler, ``Constraints on SU(5)-type leptoquarks'',
  \textit{ Phys. Lett.} \textbf{ B177} (1986) 377.
\href{http://dx.doi.org/10.1016/0370-2693(86)90771-9}{\texttt{
  doi:10.1016/0370-2693(86)90771-9}}.

\bibitem{PhysRevD.71.057503}
M.~Kr{\"a}mer {et~al.}, ``{Pair Production of Scalar Leptoquarks at the LHC}'',
  \textit{ Phys. Rev.} \textbf{ D71} (2005) 057503.
\href{http://dx.doi.org/10.1103/PhysRevD.71.057503}{\texttt{
  doi:10.1103/PhysRevD.71.057503}}.

\bibitem{Acosta}
D.~Acosta and S.~Blessing, ``Leptoquark Searches at HERA and the Tevatron'',
  \textit{ Ann. Rev. Nucl. Part. Sci.} \textbf{ 49} (1999) 389.
\href{http://dx.doi.org/10.1146/annurev.nucl.49.1.389}{\texttt{
  doi:10.1146/annurev.nucl.49.1.389}}.

\bibitem{Hewett}
L.~Hewett and T.~G. Rizzo, ``Much Ado About Leptoquarks: A Comprehensive
  Analysis'', \textit{ Phys. Rev.} \textbf{ D56} (1997) 5709.
\href{http://dx.doi.org/10.1103/PhysRevD.56.5709}{\texttt{
  doi:10.1103/PhysRevD.56.5709}}.

\bibitem{D0}
{ D0} Collaboration, ``Search for Pair Production of Second Generation Scalar
  Leptoquarks in $p\bar{p}$ Collisions at $\sqrt{s}=1.96$ TeV'', \textit{ Phys.
  Lett.} \textbf{ B671} (2009) 224.
\href{http://dx.doi.org/10.1016/j.physletb.2008.12.017}{\texttt{
  doi:10.1016/j.physletb.2008.12.017}}.

\bibitem{CMS}
{ CMS} Collaboration, ``The CMS experiment at the CERN LHC'', \textit{ JINST}
  \textbf{ 3} (2008) S08004.
\href{http://dx.doi.org/10.1088/1748-0221/3/08/S08004}{\texttt{
  doi:10.1088/1748-0221/3/08/S08004}}.

\bibitem{Sjostrand:2006za}
T.~Sj{\"o}strand {et~al.}, ``{PYTHIA 6.4 Physics and Manual}'', \textit{ JHEP}
  \textbf{ 05} (2006) 026.
  \href{http://dx.doi.org/10.1088/1126-6708/2006/05/026}{\texttt{
  doi:10.1088/1126-6708/2006/05/026}}.

\bibitem{Nadolsky:2008zw}
P.~M. Nadolsky {et~al.}, ``{Implications of CTEQ Global Analysis for Collider
  Observables}'', \textit{ Phys. Rev.} \textbf{ D78} (2008) 013004.
\href{http://dx.doi.org/10.1103/PhysRevD.78.013004}{\texttt{
  doi:10.1103/PhysRevD.78.013004}}.

\bibitem{MADGRAPH1}
F.~Maltoni and T.~Stelzer, ``MadEvent: Automatic Event Generation with
  MadGraph'', \textit{ JHEP} \textbf{ 02} (2003) 027.
  \href{http://dx.doi.org/10.1088/1126-6708/2003/02/027}{\texttt{
  doi:10.1088/1126-6708/2003/02/027}}.

\bibitem{MADGRAPH2}
J.~Alwall {et~al.}, ``{MadGraph/MadEvent} v4: The New Web Generation'',
  \textit{ JHEP} \textbf{ 09} (2007) 028.
\href{http://dx.doi.org/10.1088/1126-6708/2007/09/028}{\texttt{
  doi:10.1088/1126-6708/2007/09/028}}.

\bibitem{antikt}
M.~Cacciari, G.~Salam, and G.~Soyez, ``The anti k$_T$ jet clustering
  algorithm'', \textit{ JHEP} \textbf{ 04} (2008) 063.
\href{http://dx.doi.org/10.1088/1126-6708/2008/04/063}{\texttt{
  doi:10.1088/1126-6708/2008/04/063}}.

\bibitem{bayes1}
{ Particle Data Group} Collaboration, ``The Review of Particle Physics -
  Section 33.3.1'', \textit{ J. Phys.} \textbf{ G37} (2010) 075021.
  \href{http://dx.doi.org/10.1088/0954-3899/37/7A/075021}{\texttt{
  doi:10.1088/0954-3899/37/7A/075021}}.

\bibitem{bayes}
I.~Bertram, G.~Landsberg, J.~Linnemann{ et~al.}, ``{A Recipe for the
  construction of confidence limits}'', technical report, 2000.
\newblock
  \href{http://lss.fnal.gov/archive/test-tm/2000/fermilab-tm-2104.pdf}{FERMILA%
B-TM-2104}.

\bibitem{ttdilep}
{ {CMS}} Collaboration, {CMS Collaboration}, ``{First Measurement of the Cross
  Section for Top-Quark Pair Production in Proton-Proton Collisions at
  $\sqrt{s}=7$ TeV}''. {Submitted to \textit{Physics Letters B}}, 2010.
\href{http://www.arXiv.org/abs/1010.5994}{\texttt{ arXiv:1010.5994}}.

\bibitem{lumi}
{ CMS} Collaboration, ``Measurement of {CMS} Luminosity'', \textit{ CMS Physics
  Analysis Summary} \textbf{
  \href{http://cdsweb.cern.ch/record/1279145}{CMS-PAS-EWK-10-004}} (2010).

\bibitem{JESCorr}
{ CMS} Collaboration, ``Jet Energy Corrections Determination at 7 {TeV}'',
  \textit{ CMS Physics Analysis Summary} \textbf{
  \href{http://cdsweb.cern.ch/record/1308178}{CMS-PAS-JME-10-010}} (2010).

\bibitem{CMSWandZ}
{ {CMS}} Collaboration, {CMS Collaboration}, ``{Measurement of Inclusive $W$
  and $Z$ Cross Sections in $pp$ Collisions at $\sqrt{s}=7$ {TeV}}''. Submitted
  to the \textit{Journal of High Energy Physics}, 2010.
  \href{http://www.arXiv.org/abs/1012.2466}{\texttt{ arXiv:1012.2466}}.

\end{thebibliography}\endgroup

\cleardoublepage\appendix\section{The CMS Collaboration \label{app:collab}}\begin{sloppypar}\hyphenpenalty=5000\widowpenalty=500\clubpenalty=5000\textbf{Yerevan Physics Institute,  Yerevan,  Armenia}\\*[0pt]
V.~Khachatryan, A.M.~Sirunyan, A.~Tumasyan
\vskip\cmsinstskip
\textbf{Institut f\"{u}r Hochenergiephysik der OeAW,  Wien,  Austria}\\*[0pt]
W.~Adam, T.~Bergauer, M.~Dragicevic, J.~Er\"{o}, C.~Fabjan, M.~Friedl, R.~Fr\"{u}hwirth, V.M.~Ghete, J.~Hammer\cmsAuthorMark{1}, S.~H\"{a}nsel, C.~Hartl, M.~Hoch, N.~H\"{o}rmann, J.~Hrubec, M.~Jeitler, G.~Kasieczka, W.~Kiesenhofer, M.~Krammer, D.~Liko, I.~Mikulec, M.~Pernicka, H.~Rohringer, R.~Sch\"{o}fbeck, J.~Strauss, A.~Taurok, F.~Teischinger, W.~Waltenberger, G.~Walzel, E.~Widl, C.-E.~Wulz
% ...
\vskip\cmsinstskip
% ...
1:~~Also at CERN, European Organization for Nuclear Research, Geneva, Switzerland\\
% ...

\end{sloppypar}
\end{document}